\documentclass[12pt,a4paper]{article}
\usepackage{fullpage}
\usepackage[centertags]{amsmath}
\allowdisplaybreaks[4]
\usepackage{amsfonts}
\usepackage{amssymb}
\usepackage{bm}
\usepackage[dvips]{graphicx}
\usepackage{url}
\usepackage[dvips,hyperindex]{hyperref}
\begin{document}

\begin{flushright}
\begin{tabular}{r}
\texttt{arXiv:0707.4593}
\\
\textsf{Phys. Rev. D 77, 093002 (2008)}
\end{tabular}
\end{flushright}
\vspace{1cm}
\begin{center}
\large\bfseries
$\bm{\nu_{e}}$ disappearance in MiniBooNE
\\[0.5cm]
\normalsize\normalfont
Carlo Giunti
\\
\small\itshape
INFN, Sezione di Torino,
\\
\small\itshape
Via P. Giuria 1, I--10125 Torino, Italy
\\[0.5cm]
\normalsize\normalfont
and
\\[0.5cm]
\normalsize\normalfont
Marco Laveder
\\
\small\itshape
Dipartimento di Fisica ``G. Galilei'', Universit\`a di Padova,
\\
\small\itshape
and
\\
\small\itshape
INFN, Sezione di Padova,
\\
\small\itshape
Via F. Marzolo 8, I--35131 Padova, Italy
\end{center}
\begin{abstract}
The anomalous excess of low-energy $\nu_{e}$ events
measured in the MiniBooNE experiment is explained through
a renormalization of the absolute neutrino flux
and
a simultaneous disappearance of the $\nu_{e}$'s in the beam,
which is compatible with that indicated by the results of
Gallium radioactive source experiments.
We present the results of the fit of MiniBooNE data
($ P_{\nu_{e}\to\nu_{e}} = 0.64 {}^{+0.08}_{-0.07} $)
and
the combined fit of MiniBooNE data and the
$\nu_{e}$ disappearance measured in the Gallium radioactive source experiments,
which gives $ P_{\nu_{e}\to\nu_{e}} = 0.82 \pm 0.04 $.
We show that our interpretation of the data is also compatible with an old
indication
in favor of $\nu_{e}$ disappearance found from the analysis of
the results of beam-dump experiments,
leading to $ P_{\nu_{e}\to\nu_{e}} = 0.80 {}^{+0.03}_{-0.04} $.
\end{abstract}

\newpage

\begin{figure}[p]
\begin{center}
\includegraphics*[bb=137 255 462 590, width=\textwidth]{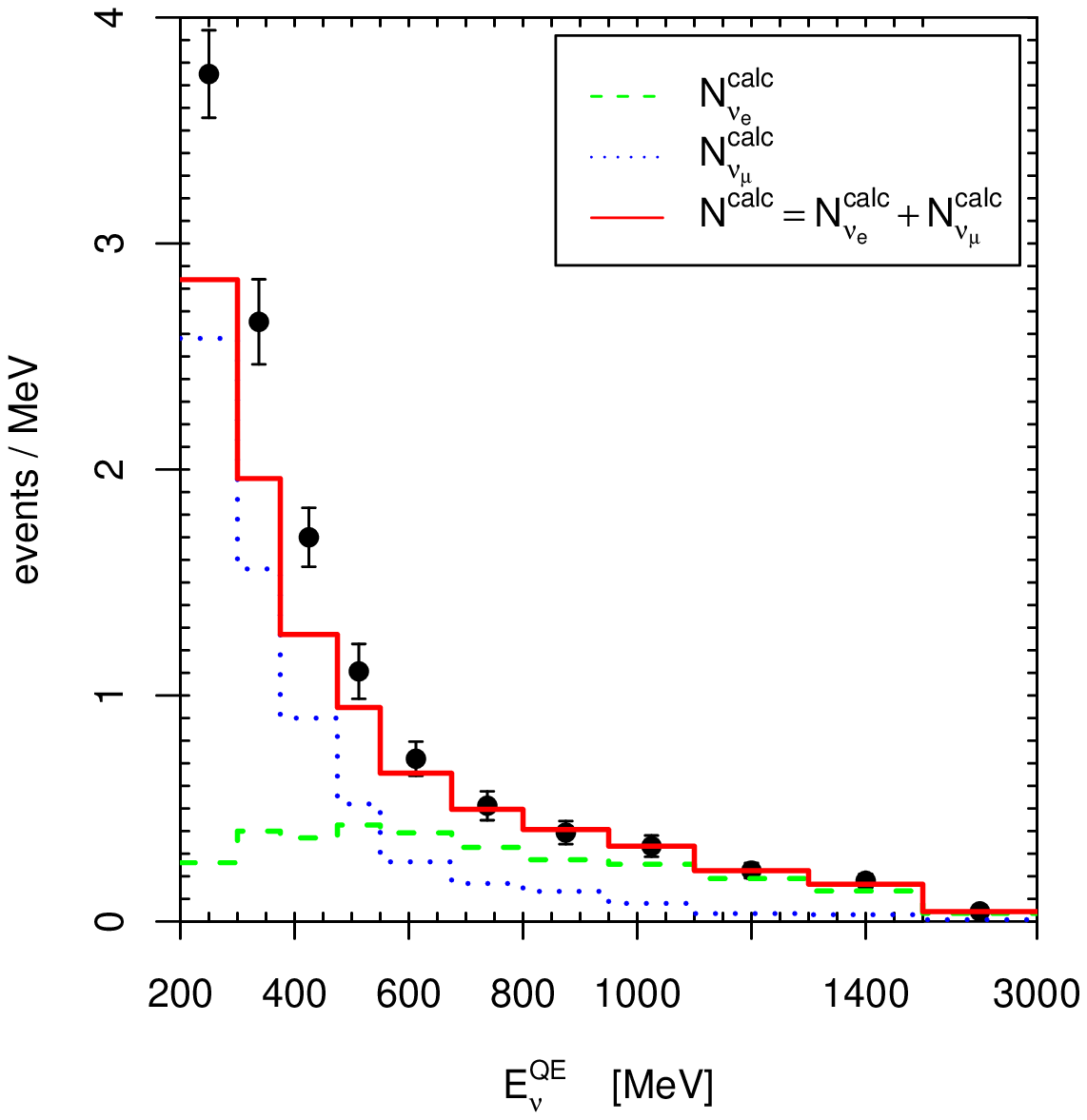}
\caption{ \label{hst-mb}
Reproduction of Fig.~2 of Ref.~\cite{0704.1500},
with the additional low-energy bin at $ 200 - 300 \, \text{MeV} $
reported in the Table in page~28 of Ref.~\cite{Tayloe-LP07}.
The points show the number of $\nu_{e}$ events measured in the
MiniBooNE experiment, with their statistical error bars.
The dashed, dotted and solid histograms show,
respectively,
the calculated number of expected
$\nu_{e}$-induced,
misidentified $\nu_{\mu}$-induced
and
total events.
}
\end{center}
\end{figure}

The MiniBooNE collaboration recently presented
\cite{0704.1500,Tayloe-LP07}
the first results of a search for $\nu_{\mu}\to\nu_{e}$ oscillations
motivated by the interpretation in terms of
$\bar\nu_{\mu}\to\bar\nu_{e}$
oscillations of the $\bar\nu_{e}$ excess observed in the LSND experiment
\cite{hep-ex/0104049}
(for reviews of the theory and phenomenology of neutrino oscillations see
Refs.~\cite{Bilenky:1978nj,Bilenky:1987ty,hep-ph/9812360,hep-ph/0202058,hep-ph/0310238,hep-ph/0405172,hep-ph/0506083,hep-ph/0606054,Giunti-Kim-2007}).
The MiniBooNE data do not show any excess of quasi-elastic charged-current
$\nu_{e}$ events with respect to the calculated background in the expected signal region,
where the reconstructed neutrino energy $ E_{\nu}^{\text{QE}} $ is larger than
$ 475 \, \text{MeV} $.
However,
the MiniBooNE data show an anomaly in the low-energy region
$ 200 \, \text{MeV} < E_{\nu}^{\text{QE}} < 475 \, \text{MeV} $,
where the $\nu_{e}$ events are significantly larger than the calculated background
(see Fig.~\ref{hst-mb}).

In this short note we discuss the possibility to explain the MiniBooNE data
through a $\nu_{e}$ disappearance which may be compatible with that indicated
by the results of the Gallium radioactive source experiments
GALLEX
\cite{Anselmann:1995ar,Hampel:1998fc}
and SAGE
\cite{Abdurashitov:1996dp,hep-ph/9803418,nucl-ex/0512041}.
In these experiments, the GALLEX and SAGE
solar neutrino detectors have been tested by placing
inside the detectors
intense artificial ${}^{51}\text{Cr}$ and ${}^{37}\text{Ar}$ sources of electron neutrinos.
The radioactive nuclei
${}^{51}\text{Cr}$ and ${}^{37}\text{Ar}$
decay through electron capture,
emitting $\nu_{e}$'s with energies
$ E_{\nu}({}^{51}\text{Cr}) \simeq 0.7 \, \text{MeV} $
and
$ E_{\nu}({}^{37}\text{Ar}) = 0.8 \, \text{MeV} $.
These neutrinos were detected through the reaction
$ \nu_{e} + {}^{71}\text{Ga} \to {}^{71}\text{Ge} + e^{-} $,
which has the low neutrino energy threshold
$ E_{\nu}^{\text{th}}({}^{71}\text{Ga}) = 0.233 \, \text{MeV} $
\cite{Kuzmin-Ga-65}.
The weighted average value of the ratio $R$ of measured and predicted ${}^{71}\text{Ge}$
production rates is
\cite{nucl-ex/0512041}
\begin{equation}
R
=
0.88 \pm 0.05
\,.
\label{s001}
\end{equation}
Since $R$ is smaller than unity by more than $2\sigma$,
it can be interpreted\footnote{
Another possible cause of $R<1$
is an overestimate of the theoretical
cross section of the Gallium detection process \cite{nucl-ex/0512041}.
}
as an indication of the disappearance of electron neutrinos
due to neutrino oscillations
\cite{Laveder:2007zz,hep-ph/0610352}.
Assuming an oscillation length
$ L_{\text{osc}} = 4 \pi E_{\nu} / |\Delta{m}^{2}| $
smaller than about 10 cm,
we obtain a relatively large squared-mass difference:
\begin{equation}
\Delta{m}^{2} \gtrsim 20 \, \text{eV}^{2}
\,.
\label{dm2}
\end{equation}

Considering now the MiniBooNE experiment,
the oscillation length corresponding to such a large $\Delta{m}^{2}$
is smaller than the
source-detector distance ($ 541 \, \text{m} $)
for neutrino energies smaller than about $ 5 \, \text{GeV} $.
Hence,
the bin-averaged survival probability of electron neutrinos $P_{\nu_{e}\to\nu_{e}}$
is practically constant in all the MiniBooNE energy bins,
whose ranges are listed in the second column of Tab.~\ref{data}.

The MiniBooNE data,
including the anomalous low-energy bins,
may be the result of a $\nu_{e}$ disappearance if the
true background is different from the calculated one by
a factor $f$ which takes into account the large uncertainties
in the calculation of the absolute normalization of
neutrino fluxes from accelerators (see Ref.\cite{physics/0609129}).
In order to understand how
$\nu_{e}$ disappearance may solve the low-energy MiniBooNE anomaly,
let us first notice that
the events which are classified as background in the MiniBooNE data analysis
are in part induced by the $\nu_{e}$'s in the beam and in part
are misidentified $\nu_{\mu}$-induced events.
The number of $\nu_{e}$-induced events ($N_{\nu_{e}}^{\text{calc}}$)
is larger than the number of misidentified $\nu_{\mu}$-induced events ($N_{\nu_{\mu}}^{\text{calc}}$)
in the high-energy bins
and smaller in the low-energy bins
(see Fig.~\ref{hst-mb}).
If the measured excess of events in the low-energy bins
is due to a larger absolute neutrino flux,
the lack of an excess in the high-energy bins can be explained by
$\nu_{e}$ disappearance,
which suppresses the event rate in the high-energy bins more than in
the low-energy bins.

\begin{figure}[p]
\begin{center}
\includegraphics*[bb=137 255 462 590, width=\textwidth]{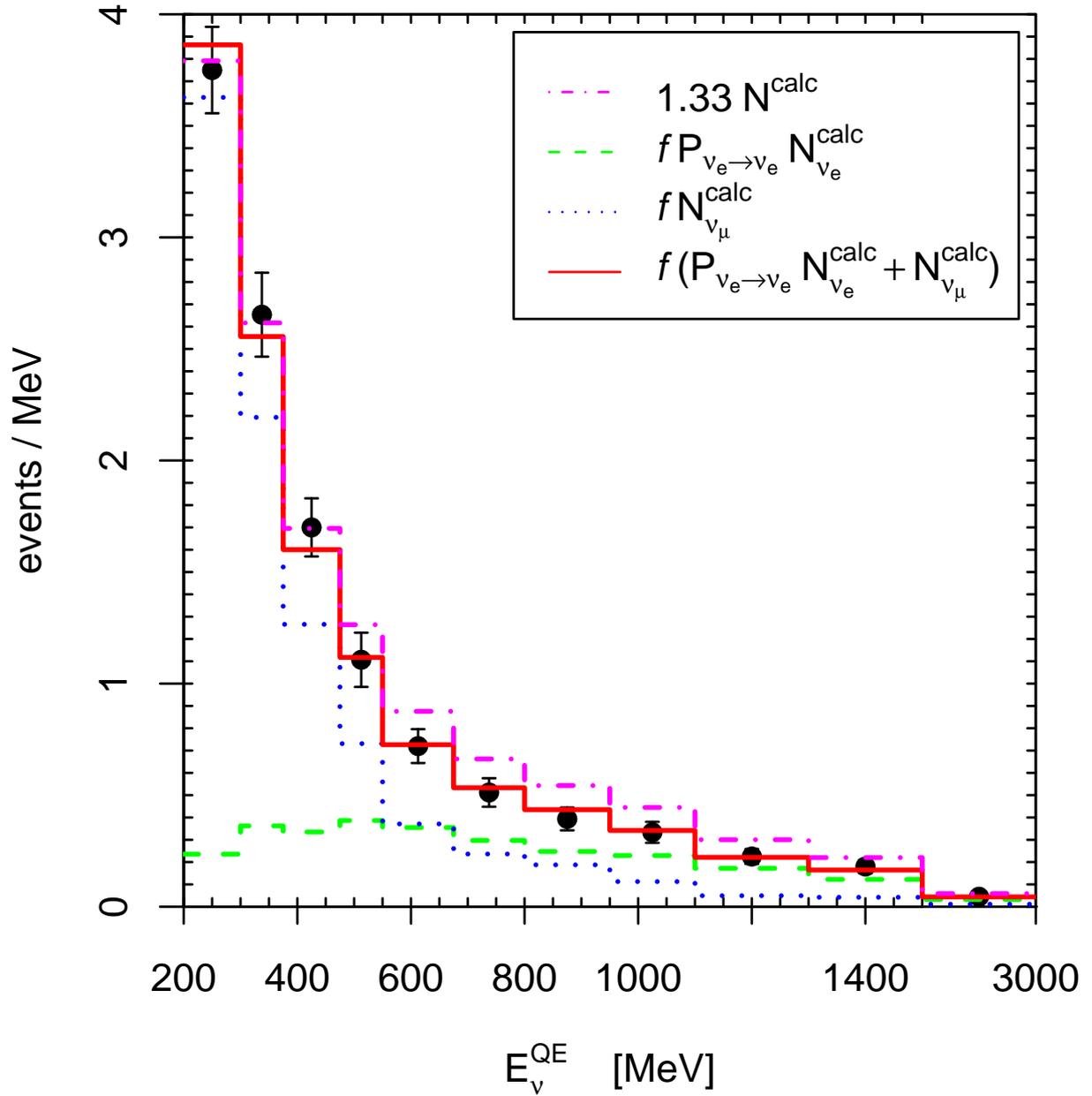}
\caption{ \label{hst-gl}
Theoretically expected number of events compared with
the MiniBooNE data,
represented by the points with their statistical error bars
(same as in Fig.~\ref{hst-mb}).
The values of $f$ and $P_{\nu_{e}\to\nu_{e}}$
are those in Eq.~(\ref{004}),
corresponding to the best fit of the MiniBooNE data.
}
\end{center}
\end{figure}

This is illustrated in Fig.~\ref{hst-gl}.
The dash-dotted histogram shows that
the three anomalous low-energy bins with
$ 200 \, \text{MeV} < E_{\nu}^{\text{QE}} < 475 \, \text{MeV} $
can be fitted by an increase of the calculated number of expected events
by a factor of about 1.33,
which is given by the ratio
$
( \sum_{j=1}^{3} N^{\text{meas}}_{j} )
/
( \sum_{j=1}^{3} N^{\text{calc}}_{j} )
$,
using the data in Tab.~\ref{data}.
However,
in this case the expected number of events in the other bins is too large.
If, instead, the expected number of $\nu_{e}$-induced events is simultaneously
suppressed by $\nu_{e}$ disappearance,
it can be kept small,
as one can see by confronting the dashed histograms in Figs.~\ref{hst-mb} and \ref{hst-gl},
which depict $ N_{\nu_{e}}^{\text{calc}} $ and $ f P_{\nu_{e}\to\nu_{e}} N_{\nu_{e}}^{\text{calc}} $,
respectively.
Then,
the high-energy bins can be fitted mainly through the
dominant small number of $\nu_{e}$-induced events.
On the other hand,
the excess in the three low-energy bins
is mainly due to the increase of the dominant
misidentified $\nu_{\mu}$-induced events
(confront the dotted histograms in Figs.~\ref{hst-mb} and \ref{hst-gl},
depicting $ N_{\nu_{\mu}}^{\text{calc}} $ and $ f N_{\nu_{\mu}}^{\text{calc}} $,
respectively).

The background calculated by the MiniBooNE collaboration
has been normalized to the measured number of charged-current quasi-elastic
$\nu_{\mu}$ events.
However, since there is an uncertainty of about 26\% \cite{Louis-Conrad-07-04-11},
a renormalization of the neutrino flux of a factor of about 1.3
cannot be excluded.

\begin{table}[t]
\begin{center}
\begin{tabular}{cccccc}
$j$
&
Energy Range [MeV]
&
$N_{\nu_{e},j}^{\text{calc}}$
&
$N_{\nu_{\mu},j}^{\text{calc}}$
&
$N^{\text{calc}}_{j}$
&
$N^{\text{meas}}_{j}$
\\
\hline
1	& $	200	-	300	$ &	26 & 258 & 284 & 375 \\
2	& $	300	-	375	$ &	30 & 117 & 147 & 199 \\
3	& $	375	-	475	$ &	37 &  90 & 127 & 170 \\
4	& $	475	-	550	$ &	32 &  39 &  71 &  83 \\
5	& $	550	-	675	$ &	49 &  33 &  82 &  90 \\
6	& $	675	-	800	$ &	41 &  21 &  62 &  64 \\
7	& $	800	-	950	$ &	41 &  20 &  61 &  59 \\
8	& $	950	-	1100	$ &	38 &  12 &  50 &  50 \\
9	& $	1100	-	1300	$ &	38 &   7 &  45 &  45 \\
10	& $	1300	-	1500	$ &	27 &   6 &  33 &  36 \\
11	& $	1500	-	3000	$ &	54 &  12 &  66 &  67 \\
\hline
\end{tabular}
\caption{ \label{data}
MiniBooNE data extracted from Fig.~2 of Ref.~\cite{0704.1500} and the Table in page~28 of Ref.~\cite{Tayloe-LP07}
(see Fig.~\ref{hst-mb}).
The six columns give:
1) bin number;
2) reconstructed neutrino energy range;
3) number of expected $\nu_{e}$-induced events (dashed histogram in Fig.~\ref{hst-mb});
4) number of expected misidentified $\nu_{\mu}$-induced events (dotted histogram in Fig.~\ref{hst-mb});
5) total number of expected events (solid histogram in Fig.~\ref{hst-mb});
6) measured number of events (points in Fig.~\ref{hst-mb}).
}
\end{center}
\end{table}

Under our hypothesis,
the theoretical number of events in the MiniBooNE $j\text{th}$ energy bin is given by
\begin{equation}
N^{\text{the}}_{j}
=
f \left(
P_{\nu_{e}\to\nu_{e}} N_{\nu_{e},j}^{\text{calc}} + N_{\nu_{\mu},j}^{\text{calc}}
\right)
\,,
\label{001}
\end{equation}
where $N_{\nu_{e},j}^{\text{calc}}$ and $N_{\nu_{\mu},j}^{\text{calc}}$
are, respectively,
the calculated number of expected $\nu_{e}$-induced and misidentified $\nu_{\mu}$-induced events
in the third and fourth columns of Tab.~\ref{data}
(corresponding to the dashed and dotted histograms in Fig.~\ref{hst-mb}).

We tested the $\nu_{e}$-disappearance hypothesis with the Pearson's chi-square
\begin{equation}
\chi^2_{\text{MB}}
=
\sum_{j=1}^{11}
\frac{ \left( N^{\text{the}}_{j} - N^{\text{meas}}_{j} \right)^2 }{ N^{\text{the}}_{j} }
\,,
\label{002}
\end{equation}
where
$N^{\text{meas}}_{j}$
are the detected events in the eleven MiniBooNE energy bins,
which are listed in the sixth column in Tab.~\ref{data}
(corresponding to the points in Figs.~\ref{hst-mb} and \ref{hst-gl}).
We found
\begin{equation}
\chi^2_{\text{MB},\text{min}} = 2.31
\,,
\label{003}
\end{equation}
with a goodness of fit of 98.6\% (9 degrees of freedom),
for
\begin{equation}
f = 1.41
\qquad
\text{and}
\qquad
P_{\nu_{e}\to\nu_{e}} = 0.64
\,.
\label{004}
\end{equation}
The solid histogram in Fig.~\ref{hst-gl} shows that these values of $f$ and $P_{\nu_{e}\to\nu_{e}}$
give an excellent fit of the data.
The increase of the expected number of
misidentified $\nu_{\mu}$-induced events
($ f N_{\nu_{\mu}}^{\text{calc}} $)
allows us to fit the three anomalous low-energy bins.
The expected number of events in the other bins is similar to that in Fig.~\ref{hst-mb},
since the increase of the expected number of
misidentified $\nu_{\mu}$-induced events is compensated
by a small decrease of the expected dominant contribution of $\nu_{e}$-induced events
($ f P_{\nu_{e}\to\nu_{e}} = 0.90 $).

\begin{figure}[p]
\begin{center}
\includegraphics*[bb=23 144 572 704, width=\textwidth]{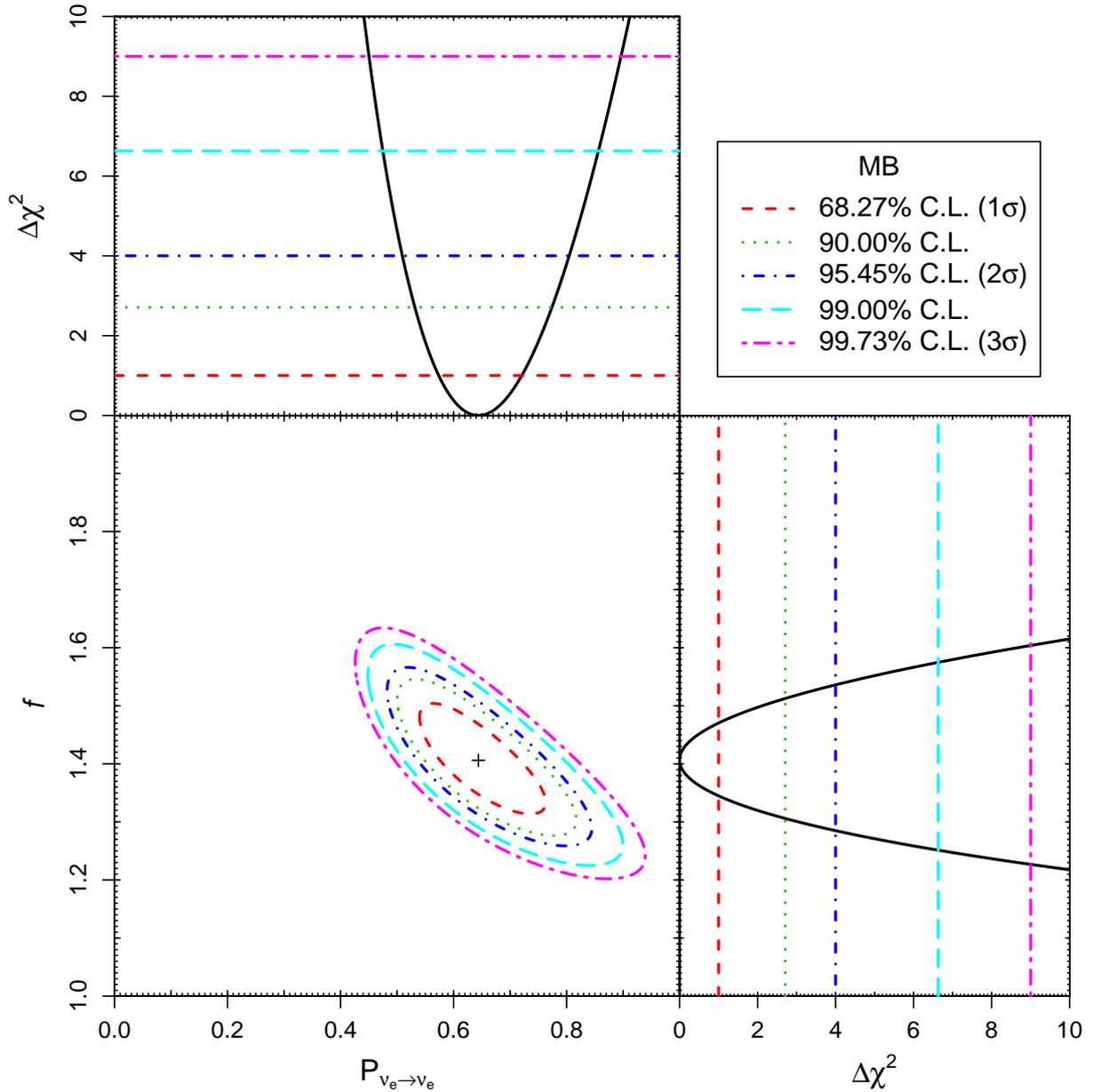}
\caption{ \label{cnt}
Allowed regions in the
$P_{\nu_{e}\to\nu_{e}}$--$f$
plane
and
marginal $\Delta\chi^2$'s
for
$P_{\nu_{e}\to\nu_{e}}$ and $f$
obtained from the fit of the MiniBooNE data.
The interrupted lines correspond to the confidence levels in the legend.
}
\end{center}
\end{figure}

The allowed regions in the
$P_{\nu_{e}\to\nu_{e}}$--$f$
plane for different confidence levels are shown in Fig.~\ref{cnt}.
One can see that there is an indication that there is indeed a
disappearance of electron neutrinos which is even larger than that
observed in the Gallium source experiments
(see Eq.~(\ref{s001})).

\begin{table}[t]
\begin{center}
\begin{tabular}{lccc}
\null \hfill C.L. \hfill \null
&
MB
&
MB+Ga
&
MB+Ga+BD
\\
\hline
Best Fit		& $0.64$ & $0.82$ & $0.80$ \\
68.27\% ($1\sigma$)	& $0.57 - 0.72$ & $0.78 - 0.86$ & $0.76 - 0.83$ \\
90.00\%			& $0.53 - 0.77$ & $0.75 - 0.89$ & $0.74 - 0.86$ \\
95.45\% ($2\sigma$)	& $0.51 - 0.80$ & $0.73 - 0.91$ & $0.73 - 0.87$ \\
99.00\%			& $0.48 - 0.86$ & $0.71 - 0.93$ & $0.71 - 0.89$ \\
99.73\% ($3\sigma$)	& $0.45 - 0.90$ & $0.69 - 0.95$ & $0.69 - 0.90$ \\
\hline
\end{tabular}
\caption{ \label{ranges-pee}
Best-fit values and allowed ranges of
$P_{\nu_{e}\to\nu_{e}}$
from the fit of MiniBooNE data (MB),
from the combined fit of MiniBooNE data and
the result in Eq.~(\ref{s001}) of Gallium radioactive source experiments
(MB+Ga)
and
from the combined fit of MiniBooNE data,
the result of Gallium radioactive source experiments
and the beam-dump indication in Eq.~(\ref{111}) of $\nu_{e}$ disappearance
(MB+Ga+BD).
}
\end{center}
\end{table}

Figure~\ref{cnt}
shows also the marginal $\Delta\chi^2$'s
for
$P_{\nu_{e}\to\nu_{e}}$
and
$f$
($ \Delta\chi^2 \equiv \chi^2 - \chi^2_{\text{min}} $).
The allowed ranges of $P_{\nu_{e}\to\nu_{e}}$
with different confidence levels are listed in Tab.~\ref{ranges-pee}.

\begin{figure}[p]
\begin{center}
\includegraphics*[bb=23 144 572 704, width=\textwidth]{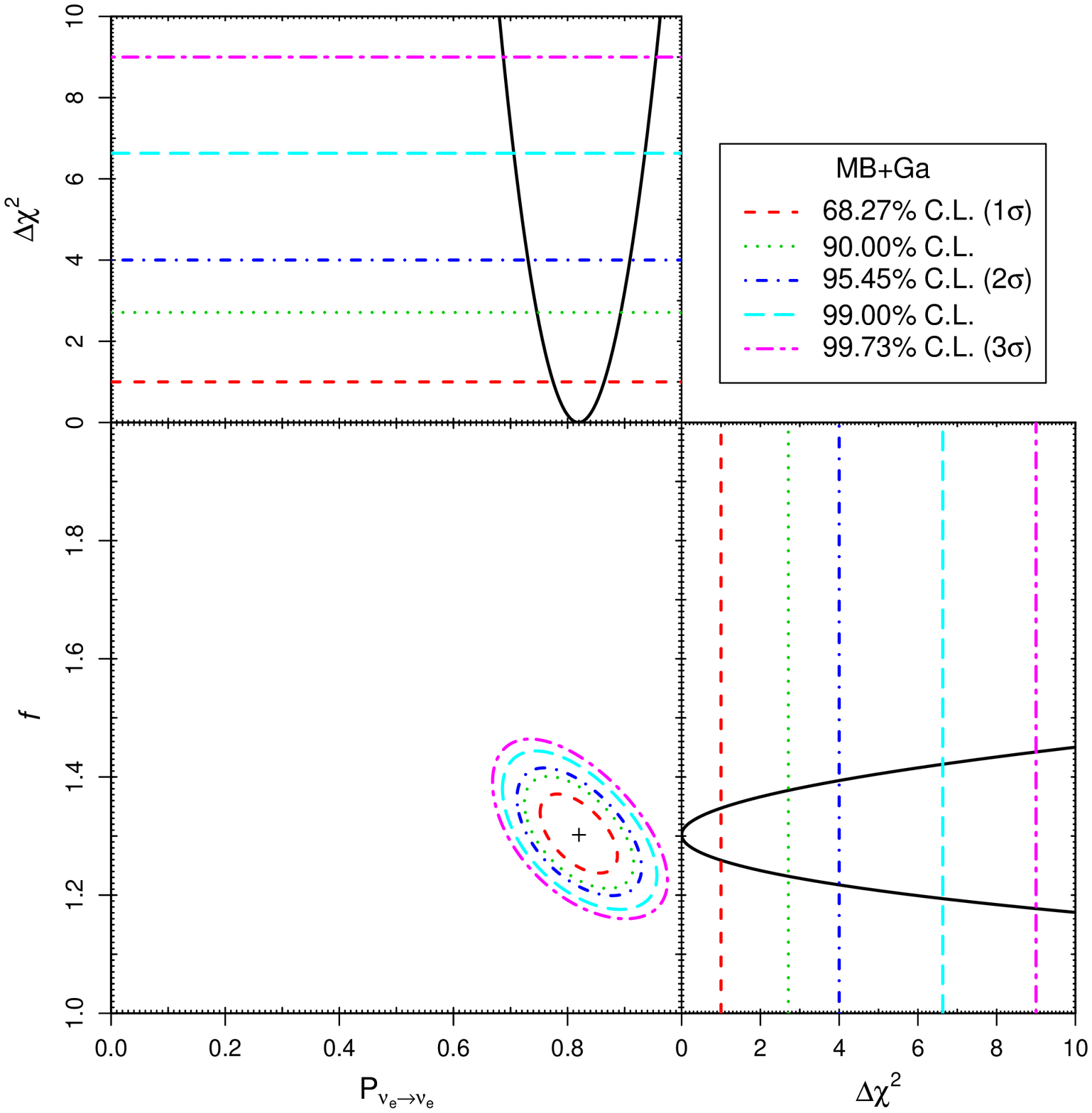}
\caption{ \label{mbga}
Allowed regions in the
$P_{\nu_{e}\to\nu_{e}}$--$f$
plane
and
marginal $\Delta\chi^2$'s
for
$P_{\nu_{e}\to\nu_{e}}$ and $f$
obtained from the combined fit of the MiniBooNE data and
the result of Gallium radioactive source experiments in Eq.~(\ref{s001}).
The interrupted lines correspond to the confidence levels in the legend.
}
\end{center}
\end{figure}

Since there is an overlap of the allowed ranges of $P_{\nu_{e}\to\nu_{e}}$
and $R$ in Eq.~(\ref{s001})
at the level of less than $2\sigma$,
we calculated the combined fit with the chi-squared
\begin{equation}
\chi^2_{\text{MB+Ga}}
=
\chi^2_{\text{MB}}
+
\left( \frac{ P_{\nu_{e}\to\nu_{e}} - 0.88 }{ 0.05 } \right)^2
\,.
\label{012}
\end{equation}
We obtained
\begin{equation}
\chi^2_{\text{MB+Ga},\text{min}} = 8.48
\,,
\label{013}
\end{equation}
with a goodness of fit of 58.2\% (10 degrees of freedom),
for
\begin{equation}
f = 1.30
\qquad
\text{and}
\qquad
P_{\nu_{e}\to\nu_{e}} = 0.82
\,.
\label{014}
\end{equation}
Figure~\ref{mbga} shows
the allowed regions in the
$P_{\nu_{e}\to\nu_{e}}$--$f$
plane for different confidence levels
and
the marginal $\Delta\chi^2$'s
for
$P_{\nu_{e}\to\nu_{e}}$
and
$f$.
The allowed ranges of $P_{\nu_{e}\to\nu_{e}}$
with different confidence levels are listed in Tab.~\ref{ranges-pee}.
Since the goodness of fit is acceptable,
the combined fit of the MiniBooNE and Gallium results provide precious information
on the value of $P_{\nu_{e}\to\nu_{e}}$
under our hypothesis for the explanation of the
MiniBooNE anomaly.
The effect of the Gallium result is to shift
the allowed regions in the
$P_{\nu_{e}\to\nu_{e}}$--$f$
plane towards
larger values of $P_{\nu_{e}\to\nu_{e}}$
and
smaller values of $f$
with respect to those obtained from the fit of the MiniBooNE data alone.

We finally consider also the old indication
in favor of $\nu_{e}$ disappearance found from the analysis of
the results of beam-dump experiments \cite{Conforto:1990sp}:
$ \sin^2 2\vartheta = 0.48 \pm 0.10 \pm 0.05 $
for the large squared-mass difference
\begin{equation}
\Delta{m}^{2}_{\text{BD}} = 377 \pm 27 \pm 7 \, \text{eV}^{2}
\,,
\label{dm2BD}
\end{equation}
which is compatible with the inequality in Eq.~(\ref{dm2}).
In this case,
the average $\nu_{e}$ survival probability is
\begin{equation}
P_{\nu_{e}\to\nu_{e}}^{\text{BD}}
=
0.76 \pm 0.06
\,.
\label{111}
\end{equation}
Notice that such a large disappearance of $\nu_{e}$
for $ \Delta{m}^{2} \sim 400 \, \text{eV}^{2} $
must be due to
transitions into sterile neutrinos,
since
$\nu_{e}\to\nu_{\mu}$ transitions are restricted by the results of the
CCFR \cite{Naples:1998va},
KARMEN \cite{Armbruster:2002mp}
and
NOMAD \cite{hep-ex/0306037}
experiments
(besides MiniBooNE itself \cite{0704.1500,Tayloe-LP07})
and
$\nu_{e}\to\nu_{\tau}$ transitions are limited by the results of the
CHORUS \cite{Eskut:2000de}
and
NOMAD \cite{Astier:2001yj}
experiments.

\begin{figure}[p]
\begin{center}
\includegraphics*[bb=23 144 572 704, width=\textwidth]{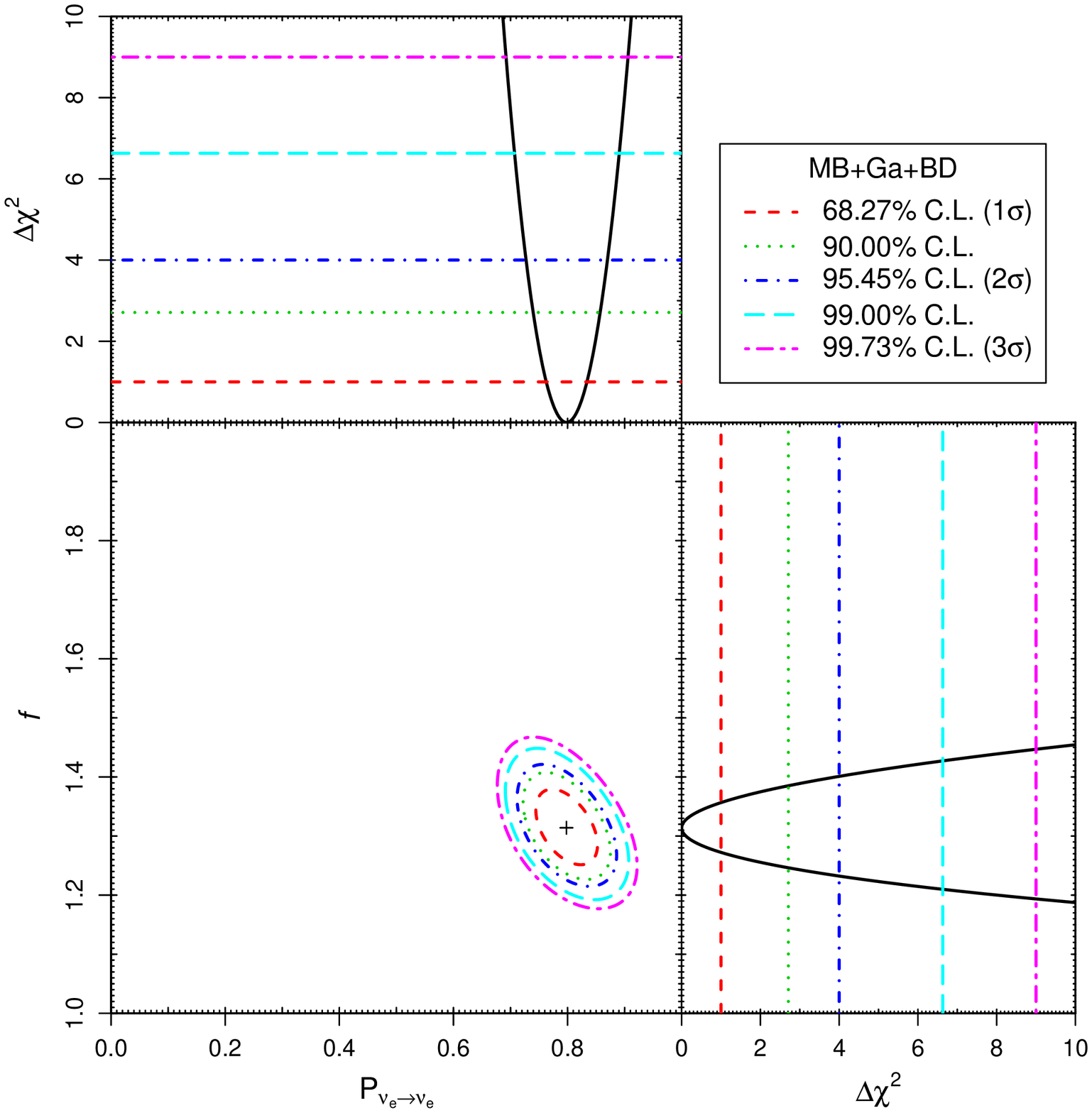}
\caption{ \label{mbgabd}
Allowed regions in the
$P_{\nu_{e}\to\nu_{e}}$--$f$
plane
and
marginal $\Delta\chi^2$'s
for
$P_{\nu_{e}\to\nu_{e}}$ and $f$
obtained from the combined fit of the MiniBooNE data,
the result of Gallium radioactive source experiments in Eq.~(\ref{s001})
and the beam-dump indication of $\nu_{e}$ disappearance in Eq.~(\ref{111}).
The interrupted lines correspond to the confidence levels in the legend.
}
\end{center}
\end{figure}

We calculated the combined fit with
the MiniBooNE data and
the result of Gallium radioactive source experiments in Eq.~(\ref{s001})
through the chi-squared
\begin{equation}
\chi^2_{\text{MB+Ga+BD}}
=
\chi^2_{\text{MB+Ga}}
+
\left( \frac{ P_{\nu_{e}\to\nu_{e}} - 0.76 }{ 0.06 } \right)^2
\,.
\label{022}
\end{equation}
We obtained
\begin{equation}
\chi^2_{\text{MB+Ga+BD},\text{min}} = 9.11
\,,
\label{023}
\end{equation}
with a goodness of fit of 61.2\% (11 degrees of freedom),
for
\begin{equation}
f = 1.31
\qquad
\text{and}
\qquad
P_{\nu_{e}\to\nu_{e}} = 0.80
\,.
\label{024}
\end{equation}
The allowed regions in the
$P_{\nu_{e}\to\nu_{e}}$--$f$
plane for different confidence levels
and
the marginal $\Delta\chi^2$'s
for
$P_{\nu_{e}\to\nu_{e}}$
and
$f$
are shown in Fig.~\ref{mbgabd}.
The allowed ranges of $P_{\nu_{e}\to\nu_{e}}$
with different confidence levels are listed in Tab.~\ref{ranges-pee}.
One can see that the allowed regions are shifted towards
slightly lower values of $P_{\nu_{e}\to\nu_{e}}$
with respect to those obtained from the fit of the MiniBooNE data and
the result of Gallium radioactive source experiments.

Notice that we assume that
the large $\Delta{m}^{2}$ in Eq.~(\ref{dm2})
does not generate
significant $ \nu_{\mu} \leftrightarrows \nu_{e} $ transitions
and
significant disappearance of $\nu_{\mu}$'s.
Therefore,
our hypothesis cannot reconcile the LSND and MiniBooNE data.
Possibilities to reconcile the LSND and MiniBooNE data
through
3+1 four-neutrino mixing \cite{hep-ph/0610352},
3+2 four-neutrino mixing \cite{0705.0107,0706.1462,0711.2018},
neutrino decay \cite{hep-ph/0505216,0707.2285},
extra-dimensions \cite{hep-ph/0504096},
mass-varying neutrinos \cite{0710.2985},
a new light gauge boson \cite{0711.1363},
and
Lorentz-violation \cite{hep-ph/0602237,hep-ph/0606154}
have been discussed in the literature.

In our explanation of the low-energy MiniBooNE anomaly,
the disappearance of $\nu_{e}$ quantified by $P_{\nu_{e}\to\nu_{e}}$
is into $\nu_{\tau}$ and/or one or more sterile neutrinos.
This is compatible with the observation of solar and reactor neutrino oscillations due to the squared-mass difference
$ \Delta{m}^{2}_{\text{SOL}} = ( 7.59 \pm 0.21 ) \times 10^{-5} \, \text{eV}^{2} $
\cite{0801.4589}
and the observation of atmospheric and accelerator neutrino oscillations due to the squared-mass difference
$ \Delta{m}^{2}_{\text{ATM}} = ( 2.74 {}^{+0.44}_{-0.26} \times 10^{-3} ) \times 10^{-3} \, \text{eV}^{2} $
\cite{Kordosky:2007gu}
if there are at least four massive neutrinos
(see Refs.~\cite{hep-ph/9812360,hep-ph/0202058,hep-ph/0606054,Giunti-Kim-2007}).
Considering the simplest case of 3+1 four-neutrino mixing with one sterile neutrino $\nu_{s}$,
the heavy neutrino $\nu_{4}$ with mass
\begin{equation}
m_{4} \simeq \sqrt{\Delta{m}^{2}} \gtrsim 4 \, \text{eV}
\label{m4}
\end{equation}
must have a very small mixing with $\nu_{\mu}$.
If the atmospheric neutrino oscillations occur in the
$\nu_{\mu}\to\nu_{\tau}$ channel,
as indicated by Super-Kamiokande data
\cite{hep-ex/0607059},
the heavy neutrino $\nu_{4}$ is mainly mixed with $\nu_{e}$ and $\nu_{s}$.
In this case,
the MiniBooNE $P_{\nu_{e}\to\nu_{e}}$ is due to
$\nu_{e}\to\nu_{s}$ transitions.

A short-baseline disappearance of electron neutrinos due to $\nu_{e}\to\nu_{s}$ transitions
affects the interpretation of the measurements of the electron neutrino flux
in all experiments with an initial $\nu_{e}$ beam.
At present,
solar and atmospheric neutrino experiments
have initial $\nu_{e}$ beams.
However, the solar neutrino data and our knowledge of the initial flux are not
sufficient to exclude an energy-independent disappearance of $\nu_{e}$'s into sterile states
at a level of about 20\%
\cite{hep-ph/0406294}.
Actually,
a comparison of the SNO Neutral-Current (NC) data
with the Standard Solar Model (SSM) prediction favors
$\nu_{e}\to\nu_{s}$ transitions
\cite{hep-ph/0610352}.
In the case of atmospheric neutrinos,
the estimated uncertainty on the initial $\nu_{e}$ flux is about 30\%
(see Ref.~\cite{Giunti-Kim-2007}).
This is too large to constrain the energy-independent $\nu_{e}\to\nu_{s}$ transitions
which explain the MiniBooNE and Gallium source experiment anomalies.

In the 3+1 four-neutrino mixing scheme discussed above with the heavy neutrino mass in Eq.~(\ref{m4})
(see Refs.~\cite{hep-ph/9812360,hep-ph/0202058,hep-ph/0606054,Giunti-Kim-2007}),
the average survival probability of electron neutrinos in the MiniBooNE experiment
is given by
\begin{equation}
P_{\nu_{e}\to\nu_{e}}
=
1 - \frac{1}{2} \, \sin^2 2 \vartheta
\,,
\label{pee}
\end{equation}
with the effective mixing angle $\vartheta$ related to the element $U_{e4}$ of the mixing matrix by
\begin{equation}
\sin^2 2 \vartheta
=
4 \, |U_{e4}|^2 \left( 1 - |U_{e4}|^2 \right)
\,.
\label{s2t}
\end{equation}
Considering the results of the fit of MiniBooNE and Gallium data in Tab.~\ref{ranges-pee},
we have
\begin{equation}
2.6 \times 10^{-2} \lesssim |U_{e4}|^2 \lesssim 0.19
\qquad
(3\sigma)
\,.
\label{ue4}
\end{equation}
Since the effective neutrino mass in tritium $\beta$-decay experiments
is given by \cite{Shrock:1980vy,McKellar:1980cn,Kobzarev:1980nk}
\begin{equation}
m_{\beta}^{2} = \sum_{k=1}^{4} |U_{ek}|^{2} \, m_{k}^{2}
\,,
\label{mbeta}
\end{equation}
from Eqs.~(\ref{m4}) and (\ref{ue4})
we have
\begin{equation}
m_{\beta} \geq |U_{e4}| \, m_{4} \gtrsim 0.7 \, \text{eV}
\,.
\label{mb}
\end{equation}
This lower bound is close to the Mainz \cite{hep-ex/0412056}
and Troitzk \cite{Lobashev:1999tp}
upper limit,
$
m_{\beta}^{(\text{exp})}
<
2.3 \, \text{eV}
\quad
(95\% \, \text{CL})
$,
and can be tested in the future KATRIN experiment \cite{hep-ex/0309007},
which will reach a sensitivity of about $0.2 \, \text{eV}$.
From Eqs.~(\ref{ue4}), (\ref{mb}) and the experimental upper limit on $m_{\beta}$
it is possible to constrain from above the value of $m_{4}$.
In order to obtain a robust upper bound we calculated the Bayesian
$3\sigma$ upper limit on $m_{\beta}$ from
$ (m_{\beta}^{(\text{exp})})^2 = -0.6 \pm 2.2 \pm 2.1 $
measured in the Mainz experiment \cite{hep-ex/0412056}:
\begin{equation}
m_{\beta}^{(\text{exp})}
<
2.9 \, \text{eV}
\qquad
(3\sigma)
\,.
\label{mbexp}
\end{equation}
From Eqs.~(\ref{ue4}), (\ref{mb}) and (\ref{mbexp}) we finally obtain
$ m_{4} \lesssim 18 \, \text{eV} $.
Hence, we have the following allowed ranges
for $m_{4}$ and $\Delta{m}^{2}$:
\begin{eqnarray}
&
4 \, \text{eV}
\lesssim
m_{4}
\lesssim
18 \, \text{eV}
&
\,,
\label{m4-range}
\\
&
20 \, \text{eV}^{2}
\lesssim
\Delta{m}^{2}
\lesssim
330 \, \text{eV}^{2}
&
\,.
\label{dm2-range}
\end{eqnarray}
Notice that the upper bound for $\Delta{m}^{2}$ in Eq.~(\ref{dm2-range})
is marginally compatible with the beam-dump value of $\Delta{m}^{2}$
in Eq.~(\ref{dm2BD}).

If massive neutrinos are Majorana particles
the amplitude of neutrinoless double-$\beta$ decay is proportional to the effective Majorana mass
\begin{equation}
m_{2\beta}
=
\left|
\sum_{k=1}^{4}
U_{ek}^2 \, m_{k}
\right|
\,.
\label{d03}
\end{equation}
From Eqs.~(\ref{ue4}) and (\ref{m4-range}) we obtain
\begin{equation}
0.1 \, \text{eV}
\lesssim
|U_{e4}|^2 \, m_{4}
\lesssim
3.4 \, \text{eV}
\,.
\label{d04}
\end{equation}
If the contributions to $m_{2\beta}$ of the other neutrino masses is much smaller,
$ m_{2\beta} \simeq |U_{e4}|^2 \, m_{4} $
is constrained in
the range (\ref{d04}),
which is compatible with the most stringent bounds
obtained in the Heidelberg-Moscow \cite{Klapdor-Kleingrothaus:2001yx} and IGEX \cite{Aalseth:2002rf} experiments
($ m_{2\beta} \lesssim 0.3 - 1.0 \, \text{eV} $; see Ref.~\cite{Giunti-Kim-2007}),
with the recent CUORICINO measurement \cite{0802.3439}
($ m_{2\beta} \lesssim 0.19 - 0.68 \, \text{eV} $)
and with the alleged\footnote{
This measurement is controversial
\cite{hep-ph/0201291,hep-ex/0202018,hep-ex/0309016}.
The issue can only be settled by future experiments
(see Ref.~\cite{hep-ph/0405078}).
}
observation of $^{76}\text{Ge}$ neutrinoless double-$\beta$ decay
due to
$ m_{2\beta} \simeq 0.2 - 0.6 \, \text{eV} $
\cite{hep-ph/0201231,hep-ph/0404088}.

If the lepton sector is symmetric under CPT transformations,
the survival probability of neutrinos and antineutrinos are equal.
In this case,
a short-baseline survival probability of electron neutrinos
smaller than about 0.95
may appear to be in contradiction with the limits obtained in reactor neutrino oscillation experiments
(see the review in Ref.~\cite{hep-ph/0107277}),
which did not observe any disappearance of electron antineutrinos with an average energy of about 4 MeV
at distances between about 10 and 1000 m from the reactor source.
Let us notice, however,
that the oscillation length of reactor neutrinos implied by
the squared-mass difference in Eq.~(\ref{dm2}) is
shorter than about $40 \, \text{cm}$.
Hence,
in reactor neutrino experiments the oscillations are seen as an averaged energy-independent suppression
of the electron antineutrino flux,
which could be revealed only with a precise calculation
of the absolute electron antineutrino flux produced in a reactor.
This calculation is rather difficult,
because of the large number
(about $10^{3}$)
of possible fragments produced in the fission of
the four isotopes
$^{235}\text{U}$,
$^{238}\text{U}$,
$^{239}\text{Pu}$, and
$^{241}\text{Pu}$,
which generate the reactor power.
Since the branching ratio and energy spectrum of some of these
fissions have not been measured,
they must be estimated with nuclear models.
Therefore,
it is possible that
the uncertainties of the calculation of
the absolute electron antineutrino flux produced in a reactor have been underestimated.
In this case,
a short-baseline $\bar\nu_{e}$ disappearance
compatible with the ranges in Tab.~\ref{ranges-pee}
may be not excluded by the results of
reactor neutrino oscillation experiments.

The impact of $\nu_{e}\to\nu_{s}$ transitions generated by
a $\Delta{m}^{2}$ in the range in Eq.~(\ref{dm2-range})
on the phenomenology of
ultra high energy neutrinos \cite{0706.0399},
very high energy atmospheric neutrinos \cite{0709.1937},
supernova physics \cite{hep-ph/0703092,0708.3337}
and
the early Universe \cite{0711.2450}
requires detailed study.

The low-energy MiniBooNE anomaly and our explanation could be tested
in the near future in the proposed MicroBooNE experiment \cite{MicroBooNE-2007}.
Other future experiments which could check
the short-baseline disappearance of electron neutrinos and antineutrinos
with high accuracy are:
Beta-Beam experiments \cite{Zucchelli:2002sa}
with a pure $\nu_{e}$ or $\bar\nu_{e}$ beam from nuclear decay
(see the reviews in Refs.~\cite{physics/0411123,hep-ph/0605033});
Neutrino Factory experiments
with a beam composed of
$\nu_{e}$ and $\bar\nu_{\mu}$,
from $\mu^{+}$ decay,
or
$\bar\nu_{e}$ and $\nu_{\mu}$,
from $\mu^{-}$ decay
(see the review in Ref.~\cite{hep-ph/0210192,physics/0411123});
experiments with a $\bar\nu_{e}$ beam
produced in recoiless nuclear decay
and detected in recoiless nuclear antineutrino capture
\cite{hep-ph/0601079};
the LENS detector
\cite{Raghavan:1997ad,LENS-2002}
with an artificial Megacurie $\nu_{e}$ source
\cite{Grieb:2006mp}.

In conclusion,
we have presented a possible explanation of the
anomalous excess of low-energy $\nu_{e}$ events
measured in the MiniBooNE experiment \cite{0704.1500,Tayloe-LP07}.
This excess may be due to a real flux of neutrinos
in the MiniBooNE beam which is larger than the calculated one.
We have shown that in this case all the neutrino energy spectrum
measured in the MiniBooNE experiment can be fitted through
a disappearance of the $\nu_{e}$'s in the beam
which is compatible with that indicated by the results of
Gallium radioactive source experiments \cite{nucl-ex/0512041}
and that indicated by
the results of beam-dump experiments \cite{Conforto:1990sp}.

\section*{Acknowledgments}

C. Giunti would like to thank the Department of Theoretical Physics of the University of Torino
for hospitality and support.

\raggedright

\end{document}